\documentclass[aps,pra,twocolumn,showpacs,amsmath,amssymb]{revtex4-1}

\usepackage{graphicx}
\usepackage{color}
\usepackage{epstopdf}
\usepackage{amsmath}
\usepackage{bm}
\usepackage{hyperref}
\usepackage{cleveref}
\usepackage{tcolorbox}
\usepackage{braket}
\usepackage[export]{adjustbox}
\usepackage[hang]{subfigure}

\begin{document}
	
	\title{
 On the infrared cutoff for dipolar droplets
 %On the infrared cut-off for beyond mean-field corrections in dipolar Bose-Einstein condensates
 }

	\author{Liang-Jun He$^{1}$}
	\author{Fabian Maucher$^{2}$}
	\author{Yong-Chang Zhang$^{1}$}
	\email{zhangyc@xjtu.edu.cn}
	\affiliation{$^1$MOE Key Laboratory for Nonequilibrium Synthesis and Modulation of Condensed Matter, Shaanxi Key Laboratory of Quantum Information and Quantum Optoelectronic Devices, School of Physics, Xi’an Jiaotong University, Xi’an 710049, People’s Republic of China \\ 
	$^2$Faculty of Mechanical, Maritime and Materials Engineering; Department of Precision and Microsystems Engineering, Delft University of Technology, 2628 CD, Delft, The Netherlands}
		
\begin{abstract}
The beyond mean-field physics due to quantum fluctuations is often described by the Lee-Huang-Yang (LHY) correction, which can be approximately written as a simple analytical expression in terms of the mean-field employing local density approximation. This model has proven to be very successful in predicting the dynamics in dipolar Bose-Einstein condensates both qualitatively and quantitatively. Yet, a small deviation between experimental results and the theoretical prediction has been observed when comparing experiment and theory of the phase boundary of a free-space quantum droplet. For this reason, we revisit the theoretical description of quantum fluctuations in dipolar quantum gases. We study alternative cutoffs, compare them to experimental results and discuss limitations. 
\end{abstract}

\maketitle

\section{INTRODUCTION}\label{introduction}
Particles in dilute dipolar Bose-Einstein condensates (BECs) interact with contact interaction via $s$-wave scattering as well as long-range dipole-dipole interaction (DDI). Because of the anisotropic nature of the DDI, the mean-field effect becomes comparable to beyond mean-field contributions related to quantum fluctuations. This leads to a range of interesting and unexpected phenomena, such as self-bound quantum droplets \cite{ferrier2016observation,schmitt2016self,baillie2016self,wachtler2016quantum,boudjemaa2018fluctuations,bottcher2019dilute,mishra2020self,smith2021quantum,bisset2021quantum,schmidt2022self,bottcher2020new,ghosh2022droplet,chomaz2023dipolar}, supersolidity and superfluidity~ \cite{baillie2018droplet,bottcher2019transient,chomaz2019longlived,tanzi2019observation,zhang2019supersolidity,norcia2021two,hertkorn2021pattern,blakie2020supersolidity,hertkorn2021supersolidity,gallemi2022superfluid,poli2023glitches,bland2022alternating,zhang2021phases,kirkby2023spin,mukherjee2023supersolid,sohmen2021birth,bland2022twodimentional,sindik2022creation,norcia2022can,zhang2023variational,smith2023supersolidity,halder2023two,ripley2023two,arazo2023self,scheiermann2023catalyzation}, excitations~\cite{natale2019excitation,hertkorn2019fate}, and quench dynamics~\cite{bland2022twodimentional,halder2022control,kadau2016observing,kirkby2023spin,bland2022alternating,zhang2021phases,mukherjee2023supersolid,sohmen2021birth,norcia2021two}. Similarly, beyond mean-field behavior has also been discussed and observed in binary BECs~\cite{petrov2015quantum,petrov2016ultradilute,cabrera2018quantum,tengstrand2019rotating,dong2022bistable,guo2021lee,cikojevic2019universality,gautam2019self,hu2020microscopic,hu2020consistent}.

The beyond mean-field corrections caused by quantum fluctuations to the mean field were first theoretically introduced by Lee, Huang, and Yang~\cite{lee1957manybody,lee1957eigenvalues}, therefore, they are referred to as Lee-Huang-Yang (LHY) correction. The LHY correction can take different forms depending on the characteristics of the underlying physics, such as the range of interactions~\cite{fischerLHY2006,lima2011quantum,lima2012beyond,petrov2015quantum,petrov2016ultradilute}, the dimensionality~\cite{petrov2015quantum,petrov2016ultradilute,lima2011quantum,lima2012beyond,boudjemaa2019two},  multi-components~\cite{bisset2021quantum,smith2021quantum}, and  gauge fields~\cite{chen2017quantum,chen2022superfluidity,chen2020angular,gu2023liquid}. By incorporating the LHY correction into the usual Gross-Pitaevskii equation (GPE), one can obtain an extended Gross-Pitaevikii equation (eGPE)~\cite{lima2011quantum,lima2012beyond,boudjemaa2015theory,aybar2019temperature,ozturk2020temperature,petrov2015quantum,petrov2016ultradilute}. The eGPE  successfully predicted the emergence of quantum droplets in binary and dipolar BECs~\cite{baillie2016self,wachtler2016quantum,schmitt2016self,boudjemaa2018fluctuations,smith2021quantum,bisset2021quantum,bottcher2019dilute,mishra2020self,schmidt2022self,petrov2015quantum,petrov2016ultradilute,cabrera2018quantum}. Moreover, the eGPE also provided a deeper understanding of the excitation spectrum~\cite{natale2019excitation,baillie2017collective,lee2018excitations,ferrier2018scissors,petter2019probing,pal2020excitations,ilg2023ground,lee2022stability,pal2022infinite,blakie2020variational,bisset2019static,hertkorn2019fate}, supersolid states~\cite{hertkorn2019fate,bottcher2020new,hertkorn2021pattern,tanzi2019observation,baillie2018droplet,smith2023supersolidity,ripley2023two,chomaz2019longlived,natale2019excitation,blakie2020supersolidity,norcia2022can,wenzel2017striped,bottcher2019transient,hertkorn2021supersolidity,schmidt2022self,gallemi2022superfluid,sanchez2023heating,arazo2023self,scheiermann2023catalyzation,zhang2023variational,zhang2019supersolidity,ghosh2022droplet,poli2023glitches,halder2023two,bland2022alternating,zhang2021phases,kirkby2023spin,mukherjee2023supersolid,sohmen2021birth,norcia2021two,bland2022twodimentional,halder2022control,chomaz2023dipolar,sindik2022creation} as well as other remarkable phenomena arising in quantum gases that cannot be understood on the mean-field level.

These experimental observations are in good agreement with the theoretical description in both binary and dipolar BECs, which demonstrates that the eGPE description is a powerful tool to describe a range of phenomena quite accurately. 
Yet, there remains a quantitative deviation between the theoretical prediction and experimental results \cite{bottcher2019dilute,cabrera2018quantum,Ota:SPP:2020}. 

With respect to quantum droplets, we can distinguish two situations: The BECs can either form a self-bound droplet state in free space or the BECs might delocalize towards ultimatively forming a plane wave, depending on the interaction strength and particle number. The critical line between these two phases predicted by the eGPE exhibits an evident shift towards either weaker repulsive short-ranged interaction or larger particle number as compared to the experimental results for dipolar BECs~\cite{bottcher2019dilute}. This mismatch between theory and experiment might be caused by the approximations used for the derivation of LHY correction, which are most importantly the local density approximation and the infrared cutoff in momentum space. 

A way to avoid having to make a cutoff is resorting to the time-dependent Hartree-Fock-Bogoliubov (TDHFB) equations rather than the eGPE. The results obtained numerically by solving the TDHFB equations feature a good agreement with experimental results at relatively small particle numbers in dipolar BECs \cite{boudjemaa2020quantum}. 
However, whereas the TDHFB equations represent a quite accurate tool to investigate the beyond mean-field physics, they are numerically expensive to solve, particularly when considering systems with more than one spatial dimension. Therefore, it remains desirable to explore simple and nearly analytical approaches to describe the effect of quantum fluctuations as well as possible. 

A numerically cheaper avenue is to reformulate the LHY correction slightly, at the cost that the cutoff becomes spatially dependent. To derive the LHY correction, usually one calculates the Bogoliubov excitation spectrum assuming that variations in the wavefunction itself are slow compared to excitations. Therefore, one can calculate the spectrum for an unmodulated state without explicit dependence on spatial dimensions and later add the spatial dependence of the wavefunction. This method is commonly referred to as local density approximation~\cite{lima2011quantum,lima2012beyond,boudjemaa2015theory,aybar2019temperature,ozturk2020temperature}. It becomes less applicable once the two scales become comparable. 

Due to the attractive part of the DDI, the excitation spectrum of dipolar BECs is complex in the low momentum regime and, thus, leads to a finite imaginary part of the LHY correction. This can be neglected to a certain extent~\cite{bisset2016ground}. Considering that a finite size of BECs inherently imposes a threshold on the low momenta of allowed excitations, for that reason, it was suggested to reformulate the LHY correction by taking a cutoff in momentum space~\cite{bisset2016ground,wachtler2016quantum,sanchez2023heating}. 

In this paper we will compare different cutoffs, their impact on the LHY correction and the subsequent shift of the droplet phase boundary and their comparison to the experimental data. 
Among the cutoffs we discuss, we choose the healing length as a natural length scale to the momentum cutoff and show that it leads to an improved agreement to the experimental data. 

The paper is organized as follows. In Sec.~\ref{formalism} A, we present the derivation of the LHY correction via Hartree-Fock-Bogoliubov (HFB) theory to provide context~\cite{griffin1996conserving,giorgini1997thermodynamics,lima2011quantum,lima2012beyond,boudjemaa2015theory,aybar2019temperature,ozturk2020temperature}. In Sec.~\ref{formalism} B, we propose a cutoff associated with the healing length to reformulate the LHY correction and compare it with other possible types of cutoffs. In Sec.~\ref{results}, we compare the theoretical result obtained by employing different cutoffs for the LHY correction to the experimental observations and furthermore discuss the contributions of low-energy discrete excitations of a self-bound droplet~\cite{baillie2017collective}. Sec.~\ref{conclusion} provides a conclusion. 

\section{Formulation of LHY correction}\label{formalism}

\subsection{LHY Correction}\label{LHY_correction}

We consider a three-dimensional ultracold quantum gas in free space, where the particles interact via short-range repulsion as well as long-range DDI as
%For dBECs, the interaction between particles contains short-range repulsion interaction and anisotropic long-range dipole-dipole interaction, we can write the interaction potential as
\begin{equation}
	V(\mathbf{x})=g\left[\delta(\mathbf{x})+\frac{3 \epsilon_{\mathrm{dd}}}{4 \pi|\mathbf{x}|^3}\left(1-3 \frac{z^2}{|\mathbf{x}|^2}\right)\right].
	\label{Vint}
\end{equation}
Here, $g=\frac{4\pi\hbar^2 a_{s}}{M}$ with $M$ being the atomic mass, and $\epsilon_{\mathrm{dd}}=\frac{a_{\mathrm{dd}}}{a_{s}}$. $a_{\rm s}$ is the $s$-wave scattering length and can be tuned via Feshbach resonances~\cite{courteille1998observation,inouye1998observation,theis2004tuning,bottcher2019dilute}, while $a_{\mathrm{dd}}=\frac{\mu_0d^2M}{12\pi\hbar^2}$ represents the dipolar length~\cite{lima2012beyond,bottcher2019dilute,schmitt2016self} with $\mu_0$ and $d$ being the vacuum permeability and the magnetic dipole moment, respectively. 
The dynamics of such a system is governed by the following Hamiltonian
\begin{equation}
\begin{split}
	\hat{H}&= \int \mathrm{d}^3 \mathbf{x} \hat{\psi}^{\dagger}(\mathbf{x}) h_0(\mathbf{x}) \hat{\psi}(\mathbf{x}) \\
	&+\frac{1}{2} \iint \mathrm{d}^3 \mathbf{x} \mathrm{d}^3 \mathbf{x}^{\prime} \hat{\psi}^{\dagger}(\mathbf{x}) \hat{\psi}^{\dagger}\left(\mathbf{x}^{\prime}\right) V\left(\mathbf{x}-\mathbf{x}^{\prime}\right) \hat{\psi}\left(\mathbf{x}^{\prime}\right) \hat{\psi}(\mathbf{x})
\end{split}
\end{equation}
where $h_{0}(\mathbf{x})=-\frac{\hbar^{2} \nabla^{2}}{2 M}-\mu$ is the single-particle Hamiltonian containing the kinetic energy and the chemical potential $\mu$, and $\hat{\psi}(\mathbf{x})$ denotes the field operator of particles.

According to the HFB theory, the field operator can be approximately expanded as $\hat{\psi}(\mathbf{x})=\Psi(\mathbf{x})+\hat{\phi}(\mathbf{x})$ with $\Psi(\mathbf{x})=\langle \hat{\psi}(\mathbf{x}) \rangle$ being the mean-field value and $\hat{\phi}(\mathbf{x})$ representing the operator of fluctuations. By substituting this expansion into the above Hamiltonian, keeping up to the third order with respect to the fluctuations operator and combining the third-order terms into the first-order term, eventually, one can obtain the following equation by letting the corresponding coefficient be equal to zero since the first-order fluctuation must vanish for the ground state~\cite{aybar2019temperature,ozturk2020temperature}
\begin{equation}	
	\begin{split}
		\big[h_0(\mathbf{x})&+\int \mathrm{d}^3\mathbf{x}^\prime V\left(\mathbf{x}-\mathbf{x}^{\prime}\right)|\Psi(\mathbf{x}^{\prime})|^2 \big]\Psi(\mathbf{x})\\
		&+\int \mathrm{d}^{3} \mathbf{x}^{\prime} V\left(\mathbf{x}-\mathbf{x}^{\prime}\right) \tilde{n}\left(\mathbf{x}^{\prime}, \mathbf{x}^{\prime}\right)\Psi\left(\mathbf{x}\right)\\
		&+\int \mathrm{d}^{3} \mathbf{x}^{\prime} V\left(\mathbf{x}-\mathbf{x}^{\prime}\right) \tilde{n}\left(\mathbf{x}^{\prime}, \mathbf{x}\right) \Psi\left(\mathbf{x}^{\prime}\right) \\
		&+\int \mathrm{d}^{3} \mathbf{x}^{\prime} V\left(\mathbf{x}-\mathbf{x}^{\prime}\right) \tilde{m}\left(\mathbf{x}^{\prime}, \mathbf{x}\right) \Psi^{*}\left(\mathbf{x}^{\prime}\right)=0
	\end{split}
	\label{GPE}
\end{equation}
i.e., a stationary eGPE. The first line corresponds to the usual mean-field GPE, while the additional terms describe the contributions of quantum fluctuations and can be reformulated into the LHY correction as discussed later. To get this eGPE, we have assumed that the dipolar gas is in the ground state with the corresponding wave function $\Psi(\mathbf{x})$, and have defined the non-condensate density $\tilde{n}\left(\mathbf{x}^{\prime}, \mathbf{x}\right)=\braket{\hat{\phi}^{\dagger}\left(\mathbf{x}^{\prime}\right) \hat{\phi}(\mathbf{x})}$ and anomalous non-condensate density $\tilde{m}\left(\mathbf{x}^{\prime}, \mathbf{x}\right)=\braket{\hat{\phi}\left(\mathbf{x}^{\prime}\right) \hat{\phi}(\mathbf{x})}$. 

Due to the dependence on the non-condensate densities, it is not yet straightforward to deal with the ground-state dipolar BECs using Eq.~(\ref{GPE}). To further simplify the above eGPE, one needs to investigate the excitation spectrum as follows. By using the Bogoliubov transformation, the fluctuations operator can be written as 
$\hat{\phi}(\mathbf{x})  =\sum_{j}\left[ u_{j}(\mathbf{x}) \hat{\alpha}_{j}-v_{j}^{*}(\mathbf{x}) \hat{\alpha}_{j}^{\dagger}\right] $, 
where $\hat{\alpha}_{j}$ ($\hat{\alpha}_{j}^{\dagger}$) is the annihilation (creation) operator of the quasiparticles and satisfy bosonic commutation relations. The Bogoliubov amplitudes are subject to the following constraint
$\int \mathrm{d}^{3} \mathbf{x}\left[u_{j}^{*}(\mathbf{x}) u_{k}(\mathbf{x})-v_{j}^{*}(\mathbf{x}) v_{k}(\mathbf{x})\right] =\delta_{jk}$
and can be determined by the Bogoliubov-de Gennes (BdG) equations as below
\begin{equation}
	\begin{split}
		\mathcal{L}_{0}& u_{j}(\mathbf{x})+\int \mathrm{d}^{3} \mathbf{x}^{\prime} V\left(\mathbf{x}-\mathbf{x}^{\prime}\right) \Psi^{*}\left(\mathbf{x}^{\prime}\right) \Psi(\mathbf{x}) u_{j}\left(\mathbf{x}^{\prime}\right) \\
		&-\int \mathrm{d}^{3} \mathbf{x}^{\prime} V\left(\mathbf{x}-\mathbf{x}^{\prime}\right) \Psi\left(\mathbf{x}^{\prime}\right) \Psi(\mathbf{x}) v_{j}\left(\mathbf{x}^{\prime}\right)=E_{j} u_{j}(\mathbf{x})\\
		\mathcal{L}_{0}& v_{j}(\mathbf{x})+\int \mathrm{d}^{3} \mathbf{x}^{\prime} V\left(\mathbf{x}-\mathbf{x}^{\prime}\right) \Psi\left(\mathbf{x}^{\prime}\right) \Psi^{*}(\mathbf{x}) v_{j}\left(\mathbf{x}^{\prime}\right)\\
		&-\int \mathrm{d}^{3} \mathbf{x}^{\prime} V\left(\mathbf{x}-\mathbf{x}^{\prime}\right) \Psi^{*}\left(\mathbf{x}^{\prime}\right) \Psi^{*}(\mathbf{x}) u_{j}\left(\mathbf{x}^{\prime}\right)=-E_{j} v_{j}(\mathbf{x}) 
	\end{split}
	\label{bdg0}
\end{equation}
with $\mathcal{L}_{0}=h_0(\mathbf{x})+\int \mathrm{d}^3\mathbf{x}^\prime V\left(\mathbf{x}-\mathbf{x}^{\prime}\right)|\Psi(\mathbf{x}^{\prime})|^2$. By incoporating the Bogoliubov amplitudes obtained from the above BdG equations into the fluctuation operator, one can readily rewrite the non-condensate densities as   
\begin{equation}
	\begin{split}
		\tilde{n}({\bf x}^{\prime},{\bf x})=&\sum_{j}\big\{v_{j}({\bf x}^{\prime})v_{j}^{*}({\bf x}) \\
		&+N_{\rm B}(E_{j})[u_{j}^{*}({\bf x}^{\prime})u_{j}({\bf x})+v_{j}({\bf x}^{\prime})v_{i}^{*}({\bf x})]\big\}\\
		\tilde{m}({\bf x}^{\prime},{\bf x})=&-\sum_{j}\big\{u_{j}({\bf x}^{\prime})v_{j}^{*}({\bf x}) \\
		&+N_{\rm B}(E_{j})[v_{j}^{*}({\bf x}^{\prime})u_{j}({\bf x})+u_{j}^{*}({\bf x}^{\prime})v_{j}({\bf x})]\big\}
	\end{split}
	\label{depletion}
\end{equation}
where we have utilized the Bose statistics property of the quasiparticles, i.e., $\braket{\hat{\alpha}_{j}^\dagger \hat{\alpha}_{k}}=\delta_{jk}N_{\rm B}(E_j)$ and $\braket{\hat{\alpha}_{j} \hat{\alpha}_{k}}=\braket{\hat{\alpha}_{j}^\dagger \hat{\alpha}_{k}^\dagger }=0 $ with $N_{\rm B}(E)=(e^{\beta E}-1)^{-1}$ and $\beta= k_{\rm B} T$. Now we arrive at the closed Eqs.~(\ref{GPE}$\sim$\ref{depletion}), which is usually referred to as the HFB theory. These equations can be solved self-consistently, however, the calculation is quite tough even for a homogeneous flat state and usually has to resort to numerics~\cite{boudjemaa2020quantum}. 

To get around the complicated computation of the above HFB equations and reach a simple description of the effect of quantum fluctuations, one viable route is to employ the local-density approximation (LDA) \cite{giorgini1997thermodynamics,lima2011quantum,lima2012beyond,aybar2019temperature,ozturk2020temperature} through the following substitutions
\begin{equation}
	\begin{aligned}
		u_{j}(\mathbf{x})\to u(\mathbf{x},\mathbf{k})e^{\mathrm{i}\mathbf{k}\cdot\mathbf{x}},& \qquad
		v_{j}(\mathbf{x})\to v(\mathbf{x},\mathbf{k})e^{\mathrm{i}\mathbf{k}\cdot\mathbf{x}}\\
		E_j\to E(\mathbf{x},\mathbf{k}),& \qquad
		\sum_j\to\int\frac{\mathrm{d}^3 \mathbf{k}}{(2\pi)^3}
		\label{substitution}
	\end{aligned}
\end{equation}
where $u(\mathbf{x},\mathbf{k})$ and $v(\mathbf{x},\mathbf{k})$ are slowly varying
functions of $\mathbf{x}$ and are subject to the constraint $|u(\mathbf{x},\mathbf{k})|^{2}-|v(\mathbf{x},\mathbf{k})|^{2}=1 $. Under such LDA, the BdG equation~(\ref{bdg0}) can be solved analytically and result in the excitation spectrum and amplitudes as follows,
\begin{equation}
		\begin{split}
	&E(\mathbf{x},\mathbf{k})={\sqrt{\varepsilon_{\mathbf{k}}(\varepsilon_{\mathbf{k}}+2n_{0}(\mathbf{x}){\tilde{V}(\mathbf{k})})}}\\
		&|v(\mathbf{x},\mathbf{k})|^{2}=\frac{\varepsilon_{\mathbf{k}}+n_{0}(\mathbf{x}){\tilde{V}}(\mathbf{k})-E(\mathbf{x},\mathbf{k})}{2E(\mathbf{x},\mathbf{k})} \\
		&u(\mathbf{x},\mathbf{k})v^{*}(\mathbf{x},\mathbf{k})=\frac{n_{0}(\mathbf{x}){\tilde{V}}(\mathbf{k})}{2E(\mathbf{x},\mathbf{k})} 	
	\end{split}
	\label{amplitude}
\end{equation}
with $\varepsilon_{\mathbf{k}}=\frac{\hbar^2 \mathbf{k}^2}{2M}$,  $n_{0}(\mathbf{x})=|\Psi(\mathbf{x})|^{2}$, and ${\tilde{V}}({\bf{k}})=g\left[1+\epsilon_{\mathrm{d d}}\left(3\cos^{2}\theta_{\bf{k}}-1\right)\right]$ being the Fourier transformation of the interaction potential Eq.~(\ref{Vint}).

One can notice that the additional terms associated with fluctuations in Eq.~(\ref{GPE}) act like a shift of the  chemical potential $\Delta \mu$ to the mean-field value $\mu$:
\begin{equation}
	\begin{split}
\Delta \mu \Psi(\mathbf{x})&=\int \mathrm{d}^{3} \mathbf{x}^{\prime} V\left(\mathbf{x}-\mathbf{x}^{\prime}\right) \tilde{n}\left(\mathbf{x}^{\prime}, \mathbf{x}\right) \Psi\left(\mathbf{x}^{\prime}\right) \\
& +\int \mathrm{d}^{3} \mathbf{x}^{\prime} V\left(\mathbf{x}-\mathbf{x}^{\prime}\right) \tilde{m}\left(\mathbf{x}^{\prime}, \mathbf{x}\right) \Psi^{*}\left(\mathbf{x}^{\prime}\right).
	\end{split}
	\label{lhy0}
	\end{equation}
Employing LDA, this chemical potential shift can be expressed as
\begin{equation}
	\begin{split}
	&\Delta\mu(\mathbf{x})=\int{\frac{\mathrm{d}^{3}{\mathbf{k}}}{(2\pi)^{3}}}{\tilde{V}}({\mathbf{k}})\bigg\{|v({\mathbf{x}},{\mathbf{k}})|^{2}-u({\bf x},{\bf k})v^{*}({\bf x},{\bf k}) \\
	&+N_{\rm B}(E)\big[|u(\mathbf{x},\mathbf{k})|^{2}+|v(\mathbf{x},\mathbf{k})|^{2}-2 u(\mathbf{x},\mathbf{k})v^{*}(\mathbf{x},\mathbf{k})\big] 
	\bigg\}
	\end{split}
	\label{deltamu}
\end{equation}

By substituting Eq.~(\ref{amplitude}) into Eq.~(\ref{deltamu}) and performing proper renormalization, we can eventually reach the following analytical LHY correction~\cite{lima2011quantum,lima2012beyond,boudjemaa2015theory,aybar2019temperature,ozturk2020temperature}
\begin{equation}
	\Delta\mu({\bf x})=\frac{32}{3}g\sqrt{\frac{a_{s}^{3}}{\pi}}\left[\mathcal{Q}_{5}+\mathcal{R}\right]|\Psi({\bf x})|^{3} 
\end{equation}
where
\begin{equation}
	\begin{split}
	{\cal Q}_{5}(\epsilon_{\mathrm{d d}};q_{c})=&\frac{1}{4\sqrt{2}}\int_{0}^{1} \bigg[(4f(u)-q_{c}^{2})\sqrt{2f(u)+q_{c}^{2}}\\
	&-3f(u)q_c+q_c^3 \bigg]f(u) \mathrm{d} u
	\end{split} \label{eq10}
\end{equation}
and $\mathcal{R}(\epsilon_{\mathrm{d d}},\tau;q_c)=\frac{3}{4\sqrt{2}}\int_{0}^{1}\mathrm{d} u\int_{q_{c}^2}^{\infty}\mathrm{d} q\frac{{q f(u)}/{\sqrt{q+2f(u)}}}{\exp\left[\sqrt{q(q+2f(u))}/\tau\right]-1} $
are associated with the quantum and thermal fluctuations, respectively, with $f(u)=1+\epsilon_{\mathrm{dd}}\left(3u^2-1\right)$ and $\tau=\frac{k_B T}{g n_0(\mathbf{x})}$. Hereafter, we will focus on the case at zero temperature, where the thermal fluctuations vanish, and thus the LHY correction is reduced to~\cite{lima2011quantum,lima2012beyond,boudjemaa2015theory,aybar2019temperature,ozturk2020temperature}
\begin{equation}
	\Delta\mu({\bf x})=\frac{32}{3}g\sqrt{\frac{a_{s}^{3}}{\pi}}{\cal Q}_{5}(\epsilon_{\mathrm{d d}};q_c)|\Psi({\bf x})|^{3} .
\end{equation}
We would like to point out that $q_c$ appearing in $\mathcal{Q}_5$ and $\mathcal{R}$ corresponds to the cutoff in momentum space, which has a significant impact on the LHY correction and will be further discussed in Sec.~\ref{cutoff_of_LHY_correction}. Without the cutoff (i.e., $q_c=0$), $\mathcal{Q}_5$ can be simply approximated by a analytical function of $1+\frac{3}{2}\epsilon_{\mathrm{dd}}^2$ by neglecting its imaginary part, which has been widely used in the research related to the effect of quantum fluctuations in quantum gases  \cite{baillie2016self,bisset2016ground,baillie2017collective,bottcher2019dilute}. 

Here, we have omitted the term related to the non-condensate density $\tilde{n}\left(\mathbf{x}^{\prime}, \mathbf{x}^{\prime}\right)$ in Eq.~\eqref{GPE}
as it is small compared to $\Delta \mu$~\cite{aybar2019temperature,boudjemaa2020quantum}. Through a similar derivation as above for $\Delta \mu$, this term can be rewritten as
\begin{equation}
	\Delta\tilde{\mu}({\bf x})=\frac{8}{3} \sqrt{\frac{a_s^3}{\pi}}\left(\mathcal{Q}_3+\mathcal{P}\right)\int \mathrm{d}^{3} \mathbf{x}^{\prime} V\left(\mathbf{x}-\mathbf{x}^{\prime}\right)|\Psi(\mathbf{x}')|^3
\end{equation}
with $\mathcal{Q}_3\left(\epsilon_{\mathrm{dd}} ; q_c\right)= \frac{1}{\sqrt{2}} \int_0^1 \mathrm{d} u \big[\left(f(u)-q_c^2\right) \sqrt{2 f(u)+q_c^2}+q_c^3\big]$ and $\mathcal{P}=  \frac{3}{\sqrt{2}} \int_0^1 \mathrm{d} u \int_{q_c^2}^{\infty} \mathrm{d}q  \frac{{(q+f(u))}/{\sqrt{q+2 f(u)}}}{\exp \left[\sqrt{q(q+2 f(u))} / \tau\right]-1}$.
At zero temperature, it can be reduced to
\begin{equation}
	\Delta\tilde{\mu}({\bf x})=\frac{8}{3} \sqrt{\frac{a_s^3}{\pi}}\mathcal{Q}_3\left(\epsilon_{\mathrm{dd}} ; q_c\right)\int \mathrm{d}^{3} \mathbf{x}^{\prime} V\left(\mathbf{x}-\mathbf{x}^{\prime}\right)|\Psi(\mathbf{x}')|^3
	\label{tildemu}
\end{equation}
In Sec. \ref{results} we numerically examine the effect of $\Delta\tilde{\mu}({\bf x})$ and show that it merely causes a tiny shift to the critical point of a self-bound droplet in free space, which justifies the omission of this term~\cite{aybar2019temperature,boudjemaa2020quantum}. 

\subsection{Cutoff in Momentum Space}\label{cutoff_of_LHY_correction}
From Eq.~(\ref{eq10}), it is clear that $\mathcal{Q}_5$ inevitably has a finite imaginary part due to the anisotropic character of the DDI in case of a vanishing cutoff $q_c=0$. 
To avoid this artefact, we can choose to simply neglect the imaginary part of $\mathcal{Q}_5$ as it is small compared to the real part~\cite{bisset2016ground}. However, this formulation of the LHY correction leads to a deviation between the theoretical and experimental results regarding the stable regime of a single quantum droplet in free space~\cite{bottcher2019dilute}. 

In case of a finite-sized quantum droplet, the long-wavelength excitation is actually suppressed, and only the excitations with the momentum beyond the inverse of size of the droplet can be supported. Hence, the finite size of the droplet inherently imposes a momentum cutoff for the integration in Eq.~(\ref{eq10}). Assuming the corresponding sizes of the dipolar droplet along the polarization direction and the transverse directions are $\sigma_z$ and $\sigma_\rho$, respectively, two different options of the cutoff have been suggested~\cite{wachtler2016quantum,bisset2016ground}:
\begin{equation}
	\begin{split}
	&k^1_c=\sqrt{k_{c,\rho}^2\sin^2\theta+k_{c,z}^2\cos^2\theta} \\
	&k^2_c=\left(\frac{\sin^2\theta}{k_{c,\rho}^2}+\frac{\cos^2\theta}{k_{c,z}^2}\right)^{-1/2}
	\end{split}
	\label{cutoff_size}
\end{equation}
Here, $k_{c,\rho}=\frac{2\pi}{\sigma_\rho}$, $k_{c,z}=\frac{2\pi}{\sigma_z}$ and $\theta$ corresponds to the angle between the momentum and the polarization directions and is spatially dependent. 
As the droplet size increases with particle number, both cutoffs vanish for $N\rightarrow \infty$. 
We will show later that the difference between these two cutoffs becomes indeed negligible and approaches the result without cutoff at large particle numbers. In addition to the above elliptical cutoffs, Ref.~\cite{aybar2019temperature} empirically proposes a spherical cutoff
\begin{equation}
	k^3_c=\frac{\pi\sqrt{2Mgn_0(\mathbf{x})}}{2\hbar}.
	\label{cutoff_spherical}
\end{equation}
%the underlying physics of which remains to be understood.

Apart from the droplet size, the healing length $\xi={\hbar}/{\sqrt{2Mn_0(\mathbf{x})|\tilde{V}(\mathbf{k})|}}$~\cite{pethick2008bose,chomaz2023dipolar}also represents a natural characteristic length scale of BECs, and provides an inherent limit to the excitation momentum~\cite{Ketterle:PRL:1999,Davidson:PRL:2002,Vernac:PRL:2012}. 
It is the length scale within which the wavefunction can ``heal" when it is pinched (or set to zero). For example, it provides a scale for the size of the vortex core. One can identify this length scale by equating the kinetic energy to the interactions. 
Considering Eq.~(\ref{amplitude}), we see that the excitation spectrum is phonon-like when $k\ll \xi^{-1}$ (i.e., $\varepsilon_{\mathbf{k}}\ll n_0(\mathbf{x})|\tilde{V}(\mathbf{k})|$) and behaves like a free particle when $k\gg \xi^{-1}$ (i.e., $\varepsilon_{\mathbf{k}}\gg n_0(\mathbf{x})|\tilde{V}(\mathbf{k})|$). 

It seems that the phonon-like modes shall be easier to be excited because of the low excitation energy and thus dominate the effect of fluctuations. Counterintuitively, it has been observed that the fluctuations associated with these low-momentum exciations are in fact dramatically suppressed by the correlated pair excitations carrying opposite momentums~\cite{Ketterle:PRL:1999,Davidson:PRL:2002}. Therefore, we anticipate that the main contribution to the LHY correction is due to free-particle like excitations and, for that reason, propose the following  alternative cutoff using the healing length: 
\begin{equation}
	k^h_c={\xi}^{-1}=\frac{1}{\hbar}\sqrt{2Mn_0(\mathbf{x})|\tilde{V}(\mathbf{k})|}.
	\label{cutoff}
\end{equation}
This implies that we neglect the contribution of the phonon-like excitation modes. It is worth to notice that $\tilde{V}(\mathbf{k})$ depends on the direction of the momenta due to the anisotropy of the dipolar interactions. 
As $\tilde{V}(\mathbf{k})$ can be negative, we have chosen the modulus of $\tilde{V}(\mathbf{k})$ to avoid the imaginary contribution to the excitation. We checked that this specific choice does not lead to a significant change as compared to setting $\tilde{V}=0$ in domains where the interaction is negative. 

Such a cutoff associated with the healing length can also be understood from the excitation spectrum of a droplet. The above discussions are based on the LDA, which leads to an entirely continuous spectrum ranging from a low-frequency phonon-like excitation domain to a high-frequency free-particle-like regime [see Eq.~(\ref{amplitude})]. However, as discussed in, e.g., Ref.~\cite{baillie2017collective}, the excitation spectrum of a single droplet is actually composed of two distinct regimes, i.e., the low-energy bound modes and the high-energy unbound modes, which approximately correspond to the aforementioned phonon-like and free-particle-like modes. 

\begin{figure}[!ht]
	\centering
	\includegraphics[width=0.9\columnwidth]{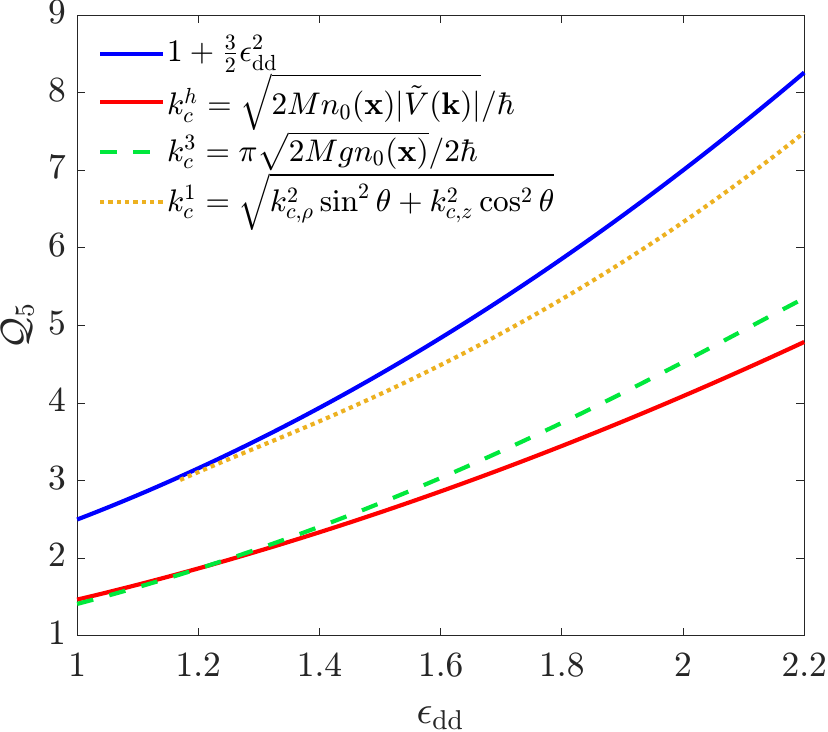}
	\caption{The coefficient of the LHY correction $\mathcal{Q}_5$ obtained using an ellipsoidal cutoff $k_c^1$ (orange), a spherical cutoff $k_c^3$ (green) and our proposed cutoff associated to healing length $k_c^h$ (red). The analytical approximation without cutoff is depicted in blue line.}
	\label{Q5}
\end{figure}

To gain an intuitive understanding of the distinction between these cutoffs, we plot the coefficient of the LHY correction as a function of $\epsilon_{\mathrm{dd}}$ (i.e., $\mathcal{Q}_5$) obtained via different options in Fig.~\ref{Q5}. The elliptical cutoffs Eq.~(\ref{cutoff_size}) are obtained by using the size of the quantum droplet close to the phase boundary. Since this type of cutoff is size-dependent, we need to first calculate the droplet using the LHY correction without cutoff and then self-consistently iteratively update the LHY correction up to convergence. As expected, the LHY correction in this case (i.e., the orange dashed line) is gradually approaching to the analytical approximation without any cutoff (i.e., the blue line) as $\epsilon_{\mathrm{dd}}$ decreases, where the critical particle number becomes large (c.f. Fig.~\ref{phase_diagram}) and thus leads to a nearly-zero cutoff. In comparison with $k_c^1$, the other elliptical cutoff $k_c^2$ leads to a smaller deviation from the analytical result (not shown)~\cite{bisset2016ground,aybar2019temperature}. In  contrast, the spherical cutoff Eq.~(\ref{cutoff_spherical}) (i.e., the green dashed line) and the cutoff induced by healing length Eq.~(\ref{cutoff}) (i.e., the red line) present a noticeable difference towards lower values of the LHY correction $\mathcal{Q}_5$.

As $\mathcal{Q}_5$ is part of the repulsive interaction, we can already foresee at this point that the droplet phase boundary will be shifted towards stronger contact interactions (i.e. larger $s$-wave scattering length) and lower particle number. Thus, an appropriate cutoff might lead to better agreement between theory and experiment~\cite{bottcher2019dilute}. Furthermore, the LHY correction obtained via our proposed cutoff converges to the result of the spherical cutoff upon decreasing $\epsilon_{\mathrm{dd}}$, while it features a clear deviance to the downside upon increasing $\epsilon_{\mathrm{dd}}$. This anisotropic deviance will also manifest itself in a slightly shifted droplet phase boundary as we will discuss in the following section. 

\section{Results and Discussions}\label{results}
To examine the reliability of our proposed cutoff, we explore the ground state phase diagram of a dipolar condensate at zero temperature in free space using the following eGPE including the LHY correction~\cite{schmitt2016self,baillie2016self,wachtler2016quantum} 
\begin{equation}
	\begin{split}
	i\hbar\frac{\partial}{\partial t}\Psi(\mathbf{x})&=\bigg[-\frac{\hbar^2\nabla^2}{2M}
	+\int V\left(\mathbf{x}-\mathbf{x}'\right)\left|\Psi(\mathbf{x}')\right|^2\mathrm{d}^3\mathbf{x}' \\
	&+\frac{32}{3}g\sqrt{\frac{a_{s}^{3}}{\pi}} \mathcal{Q}_{5}(\epsilon_{\mathrm{d d}};q_c)|\Psi({\bf x})|^{3} \bigg]\Psi(\mathbf{x}).
	\end{split}
	\label{eGPE0}
\end{equation}
Here, $N$ is the total particle number of the condensate, and the wave function has been normalized to the particle number, i.e., $\int |\Psi({\bf x})|^{2} \mathrm{d}^3\mathbf{x}=N$. 
We will evaluate the LHY correction employing the different cutoffs discussed in the last section. To identify the ground state, we numerically propagate the above eGPE using imaginary time-evolution (i.e., replacing $t$ with $-it$) and renormalize the wave function after each propagation step. 
We use cylindrical truncation for the DDI to eliminate the influence of the periodic image caused by Fourier transformation~\cite{ronen2006bogoliubov,lu2010spatial}.
Eventually the solver converges to the least damped state, unless it gets trapped at a metastable state at a local minimum of the energy. 
Our findings are illustrated in Fig.~\ref{phase_diagram}. 
We proceed with investigating the phase diagram in Sec.~\ref{phase_transition_boundary} and subsequently discuss the error related to neglecting bound parts of the spectrum in Sec.~\ref{contribution_of_the_cut_part}.

\subsection{Phase Diagram of a Single Dipolar Droplet}\label{phase_transition_boundary}
\begin{figure}[!ht]
	\centering
	\includegraphics[width=\columnwidth]{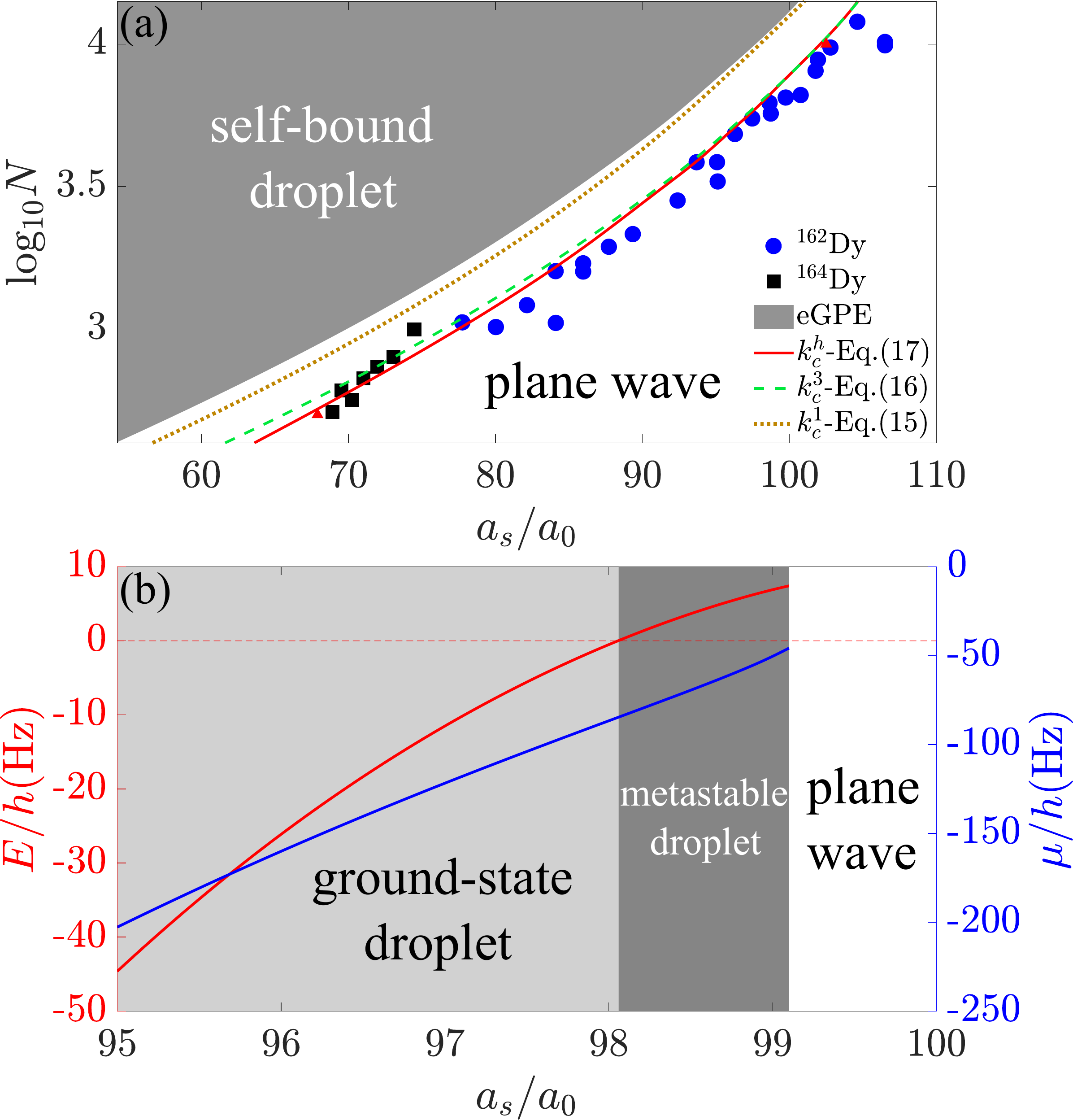}
	\caption{(a): The phase diagram of dipolar BECs in free space. The gray region indicates the self-bound droplet phase obtained with the analytical approximation of LHY correction (i.e., $\mathcal{Q}_5=1+\frac{3}{2}\epsilon_{\mathrm{dd}}^2$). The lines correspond to the transition points obtained using the ellipsoidal cutoff (orange dashed), the spherical cutoff (green dahsed), and the healing length induced cutoff (red), %and quantum Monte Carlo calculation (gray dot-dashed)~\cite{bombin2023quantum}, 
 respectively. The experimental results are illustrated by the black and blue dots \cite{bottcher2019dilute}. The red triangles depict the critical points for $N=500$ and $10000$ obtained via $k^h_c$ including $\Delta \tilde{\mu}$ [see Eq.~(\ref{tildemu})].
	(b): The energy per particle (red) and chemical potential (blue) of a droplet for $N=7000$ is presented as function of the $s$-wave scattering length. The light gray zone and dark gray zone represent regimes of a stable ground-state droplet and a droplet that is meta-stable with respect to the plane wave, respectively.}
	\label{phase_diagram}
\end{figure}

Now we would like to proceed with comparing the different approaches to the experimental data presented in Ref.~\cite{bottcher2019dilute}. 
We consider a dipolar BECs composed of $^{\text{162}}\text{Dy}$ ($a_{\mathrm{dd}}=129a_0$ with $a_0$ being the Bohr radius). 
To scan the phase boundary of the quantum droplet, we first fix the atom number $N$, relax the state via imaginary time-evolution and then gradually increase the contact interaction strength by tuning the $s$-wave scattering length $a_{s}$ up to the point where the droplet is no longer self-trapped. The results are summarized in Fig.~\ref{phase_diagram}(a).

As can be seen from Fig.~\ref{phase_diagram}(a), the stable region of the droplet (see the gray zone) predicted by the analytically approximated LHY correction without cutoff presents a deviation from the experimental observations (dots) as reported in Ref.~\cite{bottcher2019dilute}. If taking into account of the finite-size effect on the suppression of low-momentum excitations~\cite{bisset2016ground,wachtler2016quantum,sanchez2023heating} and thus using the elliptical cutoff to improve the LHY correction, the critical line is slightly shifted towards the experimental result as shown by the orange dashed line. This shift is more pronounced at small particle number, as the droplet size is smaller and thus leads to a larger finite cutoff in momentum space. However, a mismatch between theory and experiment remains. 

Let us now discuss the spherical cutoff Eq.~\eqref{cutoff_spherical} and our proposed cutoff associated with the healing length Eq.~\eqref{cutoff}. It appears that both of them are in good agreement with the experiment results (dots) as depicted by the green dashed line and the red line, respectively. 
As expected, for large particle numbers, the boundaries resulted from the healing length cutoff converges to the spherical cutoff, which is consistent with the previous analysis about the LHY coefficient in Sec.~\ref{cutoff_of_LHY_correction} (c.f. Fig.~\ref{Q5}), as the anisotropy of the wavefunction becomes small. 
Nonetheless, the difference between them grows slightly as the atom number decreases and we have to account for the anisotropy. 
Both models are in excellent with the experimental data. 
Furthermore, they remain close to what has been found via a Monte-Carlo based simulation~\cite{bombin2023quantum}, in particular for large atom numbers. 

A stable self-bound droplet can be either the ground-state (with the lowest energy) or a metastable droplet with a higher energy than a plane wave [see the shading region in Fig.~\ref{phase_diagram}(b)]~\cite{boudjemaa2020quantum}. The boundary in Fig. \ref{phase_diagram}(a) indicates the critical parameters beyond which the droplet no longer exists even as a metastable state. For illustration, the metastable region for the droplet with a particle number of $N=7000$ is shown by the dark gray zone in Fig.~\ref{phase_diagram}(b). The energy per particle becomes positive in this regime. The ground state 
in this case corresponds to a plane wave, the energy of which is fixed at zero. In contrast, the chemical potential remains negative in the metastable region until it reaches the critical point $a_{\rm s}/a_0\approx 99.1$. Such a small chemical potential implies a small number of discrete bound excitation modes around the boundary of a stable self-bound droplet~\cite{baillie2017collective}. We will discuss bound modes in the next section. 

\subsection{Contributions of the Bound Modes} \label{contribution_of_the_cut_part}

In the previous subsection, we have demonstrated that   cutoffs have a significant impact on the phase boundary.  Employing cutoffs means neglecting the contributions of all the discrete internal modes below the cutoff (i.e., the bound excitation modes of a droplet). It remains unclear whether these contributions can indeed be safely neglected. Therefore, we proceed with examining the contributions of the bound excitation modes to the LHY correction. For this purpose, we numerically calculate the Bogoliubov excitation spectrum for the self-bound droplet and then substitute them into the following equation to get the contributions of these bound modes to the LHY correction. The correction to the chemical potential can be written as 
\begin{equation}
	\Delta\mu'=-\frac{1}{\left|\Psi(\mathbf{x}) \right|^2}\sum_{j\in \mathcal{B}} \left[E_j \left|v_j(\mathbf{x})\right|^2+v_j^*(\mathbf{x})\mathcal{L}_0 v_j(\mathbf{x})\right]. 
	\label{lhy_excitation}
\end{equation}
Here, $\mathcal{B}$ refers to the set of all the bound modes. Eq.~(\ref{lhy_excitation}) is derived without using the LDA, one can find it after some algebra via combining Eqs.~(\ref{bdg0}), (\ref{depletion}), and (\ref{lhy0}). 

\begin{figure}[!b]
	\centering
	\includegraphics[width=\columnwidth]{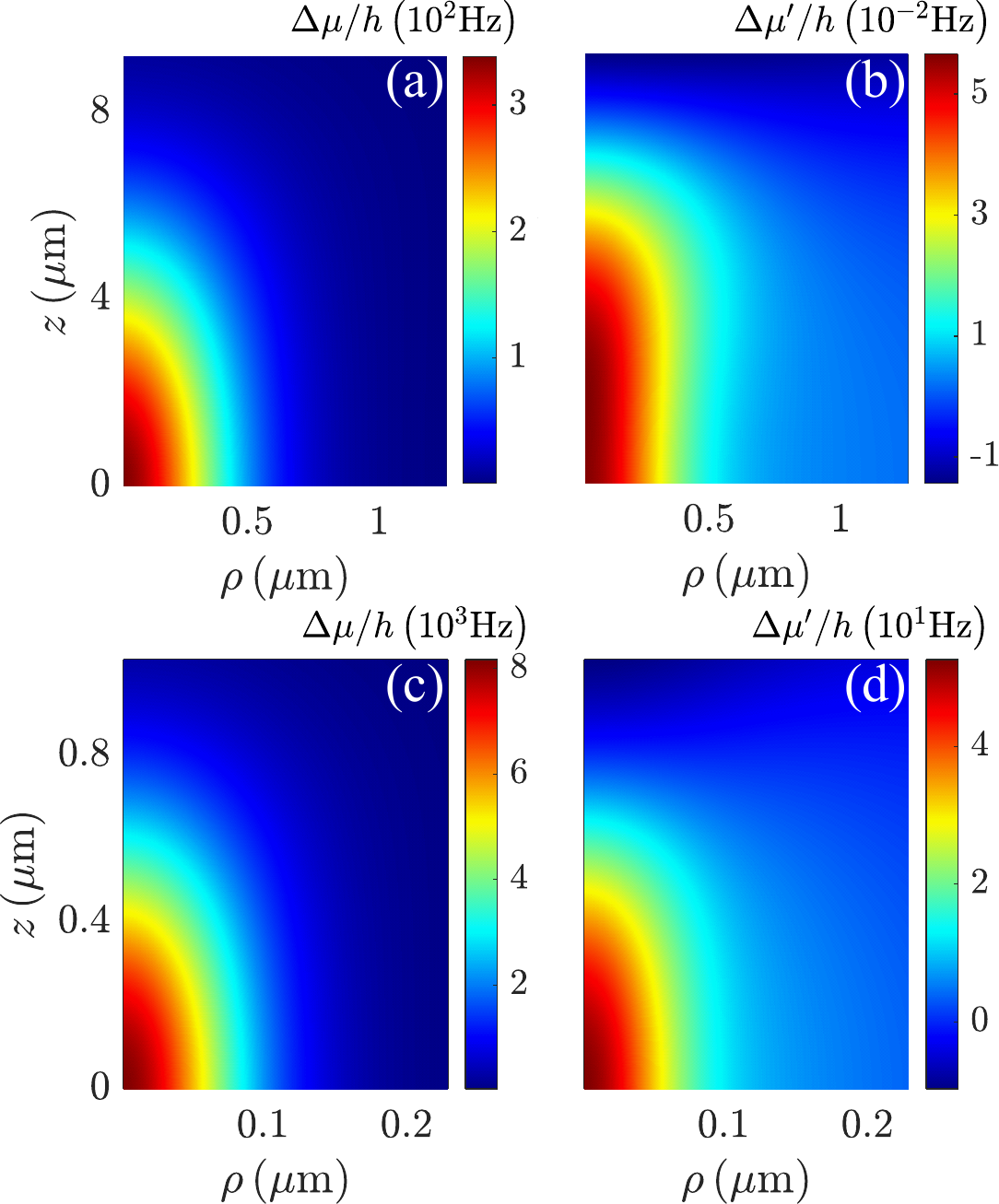}
	\caption{The contributions of free modes and bound modes to the LHY correction nearly the phase transition boundary. (a) and (c) show the contributions of continuous modes beyond the cutoff for $N=10000$ and $500$, respectively. (b) and (d) present the corresponding contributions of the discrete bound modes at $N=10000$ and $500$. For (a,b) $N=10000$, the s-wave scattering length is fixed at $a_s=102.1 a_0$, while it is equal to $67.2 a_0$ for (c,d) $N=500$.}
	\label{LHY_boundary}
\end{figure}

Fig.~\ref{LHY_boundary} shows the contributions of both free and bound excitation modes for the quantum droplets around the critical point of the phase transition. The contributions from the free modes correspond to the LHY correction obtained with healing length associated cutoff. Here we present the results for the droplets at a large particle number ($N=10000$) and at a small particle number ($N=500$), respectively. Since these states are close to the critical point, where the chemical potential approaches zero as shown in Fig. \ref{phase_diagram}(b), there is only a small numbers of the bound excitation modes, e.g., only five bounds modes for $N=10000$ and only two bound modes for $N=500$, since only bound modes with a lower energy than $-\mu$ are allowed~\cite{baillie2017collective}. As can be seen from Fig.~\ref{LHY_boundary}, the contributions of these finite bound modes are several-order smaller than the contributions of the continuous free modes. 
Such a tiny contribution of the low-energy bound excitation modes justify the cutoff for the LHY correction in the vicinity of the phase boundary. 

We have also numerically checked the influence of such discrete modes on the boundary of the self-bound droplet by adding $\Delta\mu'$ to the eGPE and computing it self-consistently during the imaginary time evolution, and found that the critical line [i.e., the red line in Fig.~\ref{phase_diagram}(a)] barely changes. Besides, we have also examined the effect of $\Delta \tilde{\mu}$ shown in Eq.~(\ref{tildemu}), which is induced by the non-condensate density $\tilde{n}(\mathbf{x}',\mathbf{x'})$ and usually neglected in previous research~\cite{aybar2019temperature,boudjemaa2020quantum}. By adding this term to the eGPE in Eq.~(\ref{eGPE0}), the critical points for the stable droplet at, e.g., $N=500$ and $10000$ have been recalculated using our proposed cutoff [see the red triangles in Fig.~\ref{phase_diagram}(a)]. In comparison with the result obtained without $\Delta \tilde{\mu}$ (i.e., the red line), the shift of the critical points is tiny and negligible. A previous investigation also shows that $\Delta \tilde{\mu}$ could be actually further reduced by higher-order correlations utilizing Beliaev formalism beyond Bogoliubov approximation~\cite{zhang2022phonon}. Hence, it is reasonable to neglect the contributions of $\Delta \tilde{\mu}$ and $\Delta \mu'$ in the eGPE, at least close to the phase boundary. 

\begin{figure}[!ht]
	\centering
	\includegraphics[width=\columnwidth]{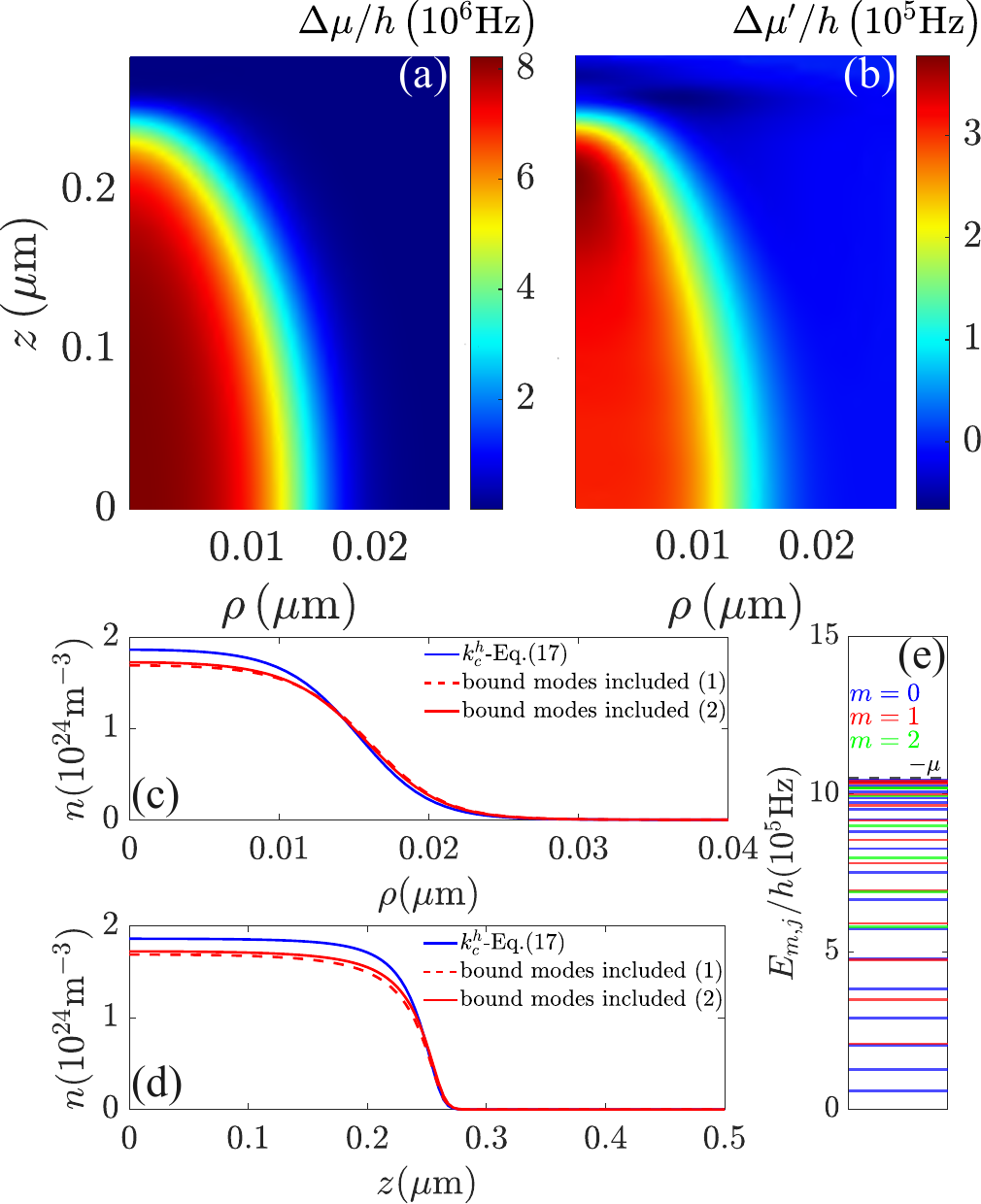}
	\caption{The LHY correction as well as the density profile of a quantum droplet state for $N=500$ at $a_s=5a_0$ (i.e., in the deeply self-bound regime). (a) and (b) present the contributions of continuous modes beyond the cutoff and that of the discrete bound modes, respectively, to the LHY correction. The density distribution of this droplet along the transverse (polarization) direction is displayed in (c) [(d)], where the blue (red) line corresponds to the ground state obtained via the LHY correction with only the continuous modes (including the discrete bound modes). The excitation spectrum of the bound modes is shown in (e).}
	\label{DeepDroplet}
\end{figure}

However, for deeply self-bound droplets far away from the critical point, the number of bound modes may grow quickly and result in a larger additional contribution to the LHY correction. For example, when $N\approx 12000$ and $a_{s}=80 a_0$, there exist more than 20 bound excitation modes~\cite{baillie2017collective}. In this case, the effect of the bound modes might no longer be negligible and possibly might need to be taken into account in Eq.~\eqref{lhy_excitation} for an accurate description. To examine the effect of these discrete bound modes, we consider the ground-state droplet in the deeply self-bound regime. Fig.~\ref{DeepDroplet} presents an example for $N=500$, where the $s$-wave scattering length is fixed at $a_s=5a_0$. In comparison with the droplet in the vicinity of the boundary [see Fig.~\ref{LHY_boundary}(c,d)], as can be seen from the Fig.~\ref{DeepDroplet}(a,b), the LHY correction is noticeably larger than that close to the boundary, since the peak density of the droplet as well as the quantity $a_{\rm dd}/a_s$ increases in the deeply self-bound region. Furthermore, it is also worth to point out that the ratio of the LHY contribution from the bound modes to that stemming from the continuous modes approximately becomes ten times larger. 
Such an increased weight of the contribution to the LHY correction is mainly due to the large number of discrete bound excitation modes, the spectrum of which is presented in Fig.~\ref{DeepDroplet}(e). To identify the ground state, we first calculate the stable solution of the eGPE~(\ref{eGPE0}) using the analytical LHY correction associated with the cutoff $k^h_c$ [see the blue line in Fig.~\ref{DeepDroplet}(c,d)]. Subsequently, we numerically address the discrete excitation spectrum of this stable solution. By adding the contribution of these bound modes to the LHY correction, we recalculate the stable solution of the updated eGPE via imaginary time evolution iteratively. Such a computation of the   spectrum computation is demanding. However, it converges already after two iterations. As shown in Fig.~\ref{DeepDroplet}(c,d), the red dashed and the red solid lines correspond to the stable solution of the first and the second iterations, respectively, presenting a good agreement with each other. Comparing the red with the blue lines, there is a visible decrease of the peak density. This is due to the fact that the discrete bound excitation modes effectively enhance the LHY correction, which behaves like a repulsive nonlinearity and thus decreases the peak density of the droplet. 

\section{CONCLUSION}\label{conclusion}
In this work, we have discussed different possible cutoffs, including a cutoff associated with the healing length. Through numerically exploring the stability of a single self-bound droplet in free space using the different cutoffs for the LHY correction, we showed that the numerical prediction presents a remarkable agreement with the previous experimental observations. Moreover, we discussed the underlying physics of this cutoff. We showed that it is related to the excitation spectrum, where the inverse healing length is a natural length scale that distinguishes between the low-energy phonon-like modes and the high-energy free-particle-like modes. 

To further quantify the approximations, we have also investigated the effect of the bound excitation modes that are neglected by this cutoff. We showed that the contribution of those discrete modes are comparably small in the vicinity of the boundary of the droplet stable region. As the number of the bound modes increases in the domain of deeply self-bound droplets at small $a_{\rm s}$, the contribution of these bound modes might become non-negligible and need to be taken into account as well. 

In comparison with the numerically demanding calculation of the HFB equations, our proposed method offers an alternative simpler route that still features high accuracy, at least close to the phase boundary. As an outlook it could also be interesting to investigate the possible cutoffs in other quantum gas systems, for example in dipolar mixtures~\cite{bisset2021quantum,smith2021quantum,boudjemaa2018fluctuations}.

\section*{Acknowledgement}

We would like to kindly thank T. Pfau for sharing the data of F. Böttcher \textit{et al.}'s work~\cite{bottcher2019dilute} we presented in Fig.~\ref{phase_diagram}(a). 
This work was supported by National Key Research and Development Program of China (Grant No.: 2021YFA1401700), the National Nature Science Foundation of China (Grant No.: 12104359), Shaanxi Academy of Fundamental Sciences (Mathematics, Physics) (Grant No.: 22JSY036). Y.C.Z. acknowledges the support of Xi'an Jiaotong University through the ``Young Top Talents Support Plan" and Basic Research Funding as well as the High-performance Computing Platform of Xi'an Jiaotong University for the computing facilities. 
%F.M. acknowledges the funding from the Ministerio de Economía y Competitividad (PID2021-128910NB-I00).

\bibliography{mybib}
\end{document}